\documentclass[twocolumn, aps, nofootinbib, prd, floats, floatfix, amsmath, amssymb, superscriptaddress, preprintnumbers]{revtex4-2} %
\usepackage{graphicx}
\usepackage{subfigure}
\usepackage{dcolumn}
\usepackage{bm}
\usepackage{amssymb}
\usepackage[utf8]{inputenc}
\usepackage[OT1]{fontenc}
\usepackage{yfonts,amsmath,amsthm,amsfonts,amssymb,amscd}
\usepackage{enumerate}
\usepackage{fancyhdr}
\usepackage{mathtools}
\usepackage{mathrsfs}
\usepackage{cancel}
\usepackage{slashed}
\usepackage{bigints}
\usepackage[flushleft]{threeparttable}
\usepackage[colorlinks=true, citecolor=purple, linkcolor=blue]{hyperref}
\usepackage{url}
\usepackage{makecell,booktabs}
\usepackage{braket}
\usepackage{relsize}
\usepackage{multirow}
\usepackage{verbatim}
\usepackage{txfonts}
\usepackage{slashed}
\usepackage{upgreek}
\usepackage{extarrows}
\usepackage{array}
\usepackage{appendix}
\usepackage[T1]{fontenc}
\usepackage{setspace}
\usepackage{aas_macros}
\usepackage[normalem]{ulem}

\usepackage[dvipsnames]{xcolor}

\usepackage{tikz} 
\usetikzlibrary{shapes,arrows,positioning,automata,backgrounds,calc,er,patterns}
\usepackage{tikz-feynman}
\tikzfeynmanset{compat=1.0.0}
\usetikzlibrary{shapes.misc}
\tikzset{cross/.style={cross out, draw=black, fill=none, minimum size=2*(#1-\pgflinewidth), inner sep=0pt, outer sep=0pt}, cross/.default={2pt}}
\usetikzlibrary{shapes.geometric}
\usepackage{orcidlink}

\begin{document}

\title{
Supermassive Primordial Black Holes from a Catalyzed Dark Phase Transition for Little Red Dots 
}

\author{Jinhui Guo}
\email{guojh23@buaa.edu.cn}
\affiliation{School of Physics, Beihang University, Beijing 100191, China}

\author{Jia Liu \orcidlink{0000-0001-7386-0253}}
\email{jialiu@pku.edu.cn}
\affiliation{School of Physics and State Key Laboratory of Nuclear Physics and Technology, Peking University, Beijing 100871, China}
\affiliation{Center for High Energy Physics, Peking University, Beijing 100871, China}

\author{Masanori Tanaka \orcidlink{0000-0002-1303-7043}}
\email{tanaka@pku.edu.cn}
\affiliation{Center for High Energy Physics, Peking University, Beijing 100871, China}

\author{Xiao-Ping Wang \orcidlink{0000-0002-2258-7741}}
\email{hcwangxiaoping@buaa.edu.cn}
\affiliation{School of Physics, Beihang University, Beijing 100191, China}

\author{Huangyu Xiao}
\email{hxiao3@bu.edu}
\affiliation{Physics Department, Boston University, Boston, MA 02215, USA}
\affiliation{Department of Physics, Harvard University, Cambridge, MA, 02138, USA}

\begin{abstract}

JWST has revealed an abundant population of compact, low-metallicity ``Little Red Dots'' (LRDs) at high redshift, challenging conventional scenarios in which supermassive black holes (SMBHs) grow from stellar-mass seeds. We consider a scenario in which the SMBHs are instead supermassive primordial black holes (SMPBHs), formed \emph{directly} in a decoupled, subdominant dark sector undergoing a first-order phase transition. Unlike conventional stochastic phase transitions, our mechanism is based on the catalysis by domain walls (DWs): most of the Universe completes the transition rapidly, while rare long-lived false-vacuum domains survive because of DW statistics and collapse into PBHs. This mechanism naturally yields SMPBH seeds with masses up to $M_{\rm PBH}\sim \mathcal{O}(10^{10}) M_\odot$, whose abundance can account for the observed LRD population. It also avoids the usual tensions with phase transition completion, $\Delta N_{\rm eff}$, and large curvature perturbations. The dark phase transition simultaneously generates an ultra-low-frequency stochastic gravitational-wave background peaking near the pulsar-timing-array range, providing a test of this dark-sector origin of LRDs.

\end{abstract}
\maketitle

\section{Introduction}
The rapid emergence of supermassive black holes (SMBHs) at high redshift remains a long-standing puzzle. Observations of luminous quasars indicate that black holes (BHs) with masses $\gtrsim 10^9\,M_\odot$ were already in place by $z \gtrsim 6$--7, i.e.\ within $\sim 1$~Gyr of the Big Bang~\cite{2006NewAR..50..665F, Banados:2017unc, Mortlock2011,Venemans2013,Wu2015,Mazzucchelli2017,Banados2018,Matsuoka2019,Yang2020,Wang2021}. Explaining such rapid growth typically requires either sustained near-/super-Eddington accretion or heavy BH seeds~\cite{Inayoshi:2019fun,2021NatRP...3..732V,Madau2001,Bromm:2002hb,Schneider2002,Koushiappas:2003zn,Begelman:2006db,Lodato:2006hw,Volonteri2008,Ferrara:2014wua,Pacucci:2015rwa,Madau:2014pta}. More recently, the James Webb Space Telescope (JWST) has revealed a large population of compact, extremely red sources at $z \gtrsim 4$, the so-called ``Little Red Dots'' (LRDs)~\cite{Matthee:2023utn,Kocevski2023,Greene2024}. These systems appear to be overmassive relative to their stellar hosts~\cite{Pacucci2023,Dayal:2024zwq,2025arXiv250821748J}, in some cases with $M_{\rm BH}\sim 10^{7}\,M_\odot$, and may contain nearly pristine gas with metallicity far below solar~\cite{Maiolino:2025tih}. Recent studies suggest that some of the unusual spectral properties of LRDs can be understood within accreting black holes~\cite{Zhang:2025bwh,Sun:2026snb,Madau:2026ekq}. The combination of large masses, large abundances, and low metallicities in some LRD systems raises questions about their early growth and assembly within standard BH seeding scenarios.

These open questions have motivated a variety of nonstandard SMBH-seeding scenarios. Examples include SMBH formation from gravothermal collapse in dissipative self-interacting dark matter halos~\cite{Pollack:2014rja,Hu:2005cd,DAmico:2017lqj,Latif:2018kqv,Padilla:2020sjy,Feng:2021rst,Xiao2021,Feng:2020kxv,Xiao2021,Lu:2024zwa,Buckley:2024eoe,Shen:2025evo}, from ultralight dark matter~\cite{Davoudiasl:2021ijv,Sikivie:2024ffa,Jiao:2025kpn}, and from primordial black hole (PBH) models in which PBHs act as massive seeds, either in clustered PBH environments~\cite{Zhang:2025tgm} or by catalyzing direct-collapse BHs and SMBH binaries that resemble LRDs~\cite{Zhang:2025grn}. In particular, recent cosmological simulations of Abell~2744--QSO1 and related LRDs suggest that their properties can be reproduced with a PBH seed of mass $M_{\rm PBH}\simeq 5\times 10^{7}\,M_\odot$~\cite{Zhang:2025oyl}.

While the PBH scenario provides an appealing solution to these observational puzzles, the \emph{direct} formation of supermassive primordial black holes (SMPBHs) is difficult to realize in realistic models while remaining consistent with observational constraints. PBHs formed from enhanced primordial curvature perturbations---aimed at $M_{\rm PBH}\sim 10^{7\text{--}9}\,M_\odot$---are typically in strong tension with CMB spectral distortion and structure formation constraints~\cite{Liu:2022lvz, Elor:2023xbz, Sui:2025epg, Xu:2025zsv, Franciolini:2025ztf,Greene:2024xgq}, unless the required fluctuations are extremely non-Gaussian or finely tuned~\cite{Carr:2020gox,Hooper:2023nnl}. A first-order phase transition (FOPT)~\cite{Kodama:1982sf, Maeda:1981gw, Blau:1986cw, Jedamzik:1999am, Garriga:2015fdk,Liu:2021svg, Hashino:2021qoq, Baker:2021nyl,Kawana:2022olo,Hashino:2022tcs, Gouttenoire:2023naa,Lewicki:2023ioy, Gouttenoire:2023gbn,Gouttenoire:2023pxh,Conaci:2024tlc,Balaji:2024rvo,Flores:2024lng,Kanemura:2024pae,Cai:2024nln,Goncalves:2024vkj,Arteaga:2024vde,Banerjee:2024cwv,Lewicki:2024sfw,Hashino:2025fse, Zou:2025sow,Franciolini:2025ztf,Balaji:2025tun, Kierkla:2025vwp,Huang:2025hos} after Big Bang nucleosynthesis (BBN) can form SMPBHs in the relevant mass range. However, in conventional scenarios, even when FOPTs occur in a subdominant dark sector, the parameter space yielding a sufficient PBH abundance is excluded by the effective number of relativistic neutrino species, $N_{\rm eff}$~\cite{An:2026hiq}.

\begin{figure}[tb]
    \centering
    \includegraphics[width=0.85\linewidth]{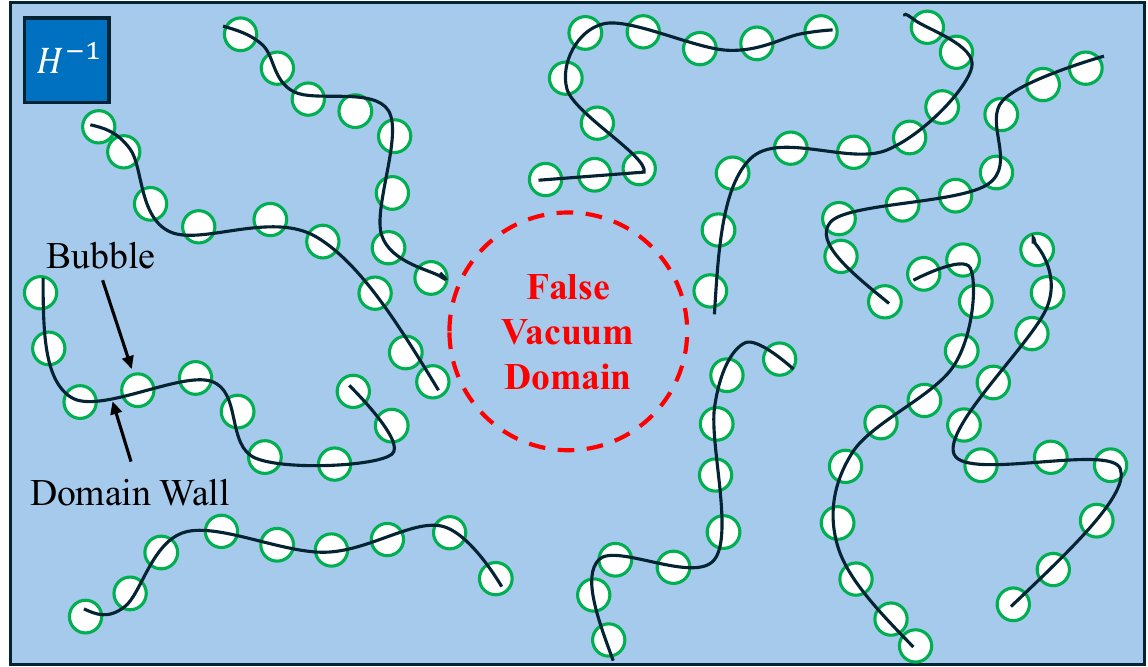}
    \caption{Schematic of a DW-catalyzed phase transition. Black curves denote DWs, and green circles denote nucleation bubbles. The red dashed circle denotes a FVD not penetrated by any DW, where no bubble nucleation occurs.}
    \label{fig:schematic-diagram}
\end{figure}

In this work, we propose PBH formation from a dark-sector phase transition (PT) catalyzed by domain walls (DWs) as an origin of the $\mathcal{O}(10^{7})\,M_\odot$ PBH seeds invoked in Ref.~\cite{Zhang:2025oyl}, and more generally of the SMPBHs associated with LRD-like systems. As shown in Fig.~\ref{fig:schematic-diagram}, in our scenario most of the Universe completes the FOPT rapidly because of catalysis by super-horizon DWs, while rare long-lived false-vacuum domains (FVDs) survive because of the spatial randomness of the DW network. When such an FVD exceeds a critical size, it collapses into a PBH~\cite{Garriga:2015fdk, Deng:2017uwc, Deng:2020mds, Gouttenoire:2023gbn, Hashino:2025fse, Ning:2026nfs}. Since PBH production is controlled by the statistics of the DW distribution, our mechanism does not generate significant curvature perturbations, and the CMB spectral distortion argument constraining \emph{direct} SMPBH formation from large primordial perturbations~\cite{DeLuca:2025nao} does not apply in the same way. In addition, the FOPT can generate a stochastic gravitational-wave background (SGWB) peaking near the pulsar timing array (PTA) band, providing a promising observational signature of this mechanism.

\section{Dark phase transition}
In our scenario, the Universe contains the SM plasma and a \emph{decoupled} dark sector, which is always subdominant in energy density. The total energy density is written as $\rho_{\rm tot} = \rho_{\rm R} + \rho_{\rm DR} + \rho_{\rm V}$, where $\rho_{\rm R}$, $\rho_{\rm DR}$ and $\rho_{\rm V}$ denote the SM radiation, dark radiation (DR), and dark vacuum energy densities, respectively. We assume $\rho_{\rm R} \gg \rho_{\rm V} \gg \rho_{\rm DR}$, so that $\rho_{\rm tot} \simeq \rho_{\rm R}$ and the cosmic expansion follows the standard radiation-dominated evolution. The small dark-to-visible energy ratio (defined as $\rho_{\rm V}/\rho_{\rm R}=\alpha$) is important for evading experimental constraints, as shown later.

Before the PT during radiation domination, both the SM radiation and DR redshift as $a^{-4}$, so that the ratio $\rho_{\rm DR}/\rho_{\rm R}$ remains constant, while $\rho_{\rm V}$ stays constant. The dark phase transition starts when the bubble nucleation rate $\Gamma(T)$ becomes comparable to the Hubble expansion rate. The corresponding temperature (time) is denoted by $T_n$ ($t_n$), and quantities with subscript ``$n$'' are evaluated at this nucleation temperature. Since the SM plasma is decoupled from the dark sector, the latent heat released during the transition is entirely injected into the DR. For a rapid dark-sector PT, we have $\rho_{\rm DR}(t_{\rm PT})\simeq \rho_{\rm V}(t_n)=\Delta V$, where $\Delta V$ is the potential difference between the false and true vacua in the dark sector, and $t_{\rm PT}$ represents the PT completion time.

\section{Cosmological constraints}
Cosmological constraints on a dark-sector PT at sub-MeV $T_{n}$ arise mainly from gravitationally induced curvature perturbations and the associated dark radiation contribution to $\Delta N_{\rm eff}$. 
We describe the implications of these constraints in our scenario. 

Since the dark sector has no interactions with the SM, the dark-sector PT does not directly inject energy into the visible plasma. 
The unavoidable imprint instead comes from a curvature perturbation $\mathcal{P}_{\zeta}(k)$ due to stochastic bubble nucleation. 
This curvature perturbation induces the $\mu$-type distortion~\cite{Chluba:2012we}, which is expressed as~\cite{Liu:2022lvz, Elor:2023xbz, Cai:2024nln, Greene:2026gnw}
\begin{equation}\label{eq:mu-distortion}
\begin{aligned}
& \mu  \simeq 
\int d\ln k~
\mathcal{P}_\zeta(k)
\left[
e^{-k/(5400\,{\rm Mpc})}
-
e^{-\left\{ k/(31.6\,{\rm Mpc}) \right\}^2}
\right], \\
&\mathcal{P}_{\zeta}(k) = \alpha^2 f(t, \alpha, \beta) \mathcal{P}_{\delta t}(k) \,, 
\end{aligned}
\end{equation}
where $f(t, \alpha, \beta)$ is a time- and model-dependent function in terms of $\alpha$ and the duration of FOPTs $\beta^{-1}$. 
$\mathcal{P}_{\delta t}(k)$ represents the dimensionless power spectrum, which is a two-point correlation function describing the time difference of bubble formation at different points.
For sub-MeV scale FOPTs, $\mu$ distortion data imposes a strong constraint~\cite{Chluba:2012we}.
We use the formulas summarized in Ref.~\cite{Greene:2026gnw} and check constraints from the FIRAS limit $\mu \lesssim 4.7 \times10^{-5}$ at the 95\% C.L.~\cite{Fixsen:1996nj, Bianchini:2022dqh}.

After the dark-sector PT, the released vacuum energy is converted into relativistic hidden-sector excitations, contributing to~\cite{Davoudiasl:2021ijv} 
\begin{equation}
\begin{aligned}
\label{eq:Neff_Tn}
    \Delta N_{\rm eff}(t_{\rm PT})=\frac{8}{7}\left(\frac{11}{4}\right)^{4/3}\frac{\rho_{\rm DR}(t_{\rm PT})}{\rho_R(t_{\rm PT})}
    \simeq 4.4\cdot\frac{\rho_{\rm DR}(t_{\rm PT})}{\rho_R(t_{\rm PT})} \,.
\end{aligned}
\end{equation} 

The current CMB bounds are $|\Delta N_{\rm eff}|\lesssim 0.3$~\cite{Planck:2018vyg}$;$ recently, Ref.~\cite{Yeh:2026pil} reported $\Delta N_{\rm eff}\lesssim 0.125$ at 95\% CL. The future expected sensitivity is $|\Delta N_{\rm eff}|\lesssim 0.06$~\cite{Abazajian:2019eic}$,$ while BBN gives $|\Delta N_{\rm eff}|\lesssim 0.4$~\cite{Pitrou:2018cgg}. Here we adopt $\Delta N_{\rm eff}\lesssim 0.3$ as the conventional bound.
In general, the vacuum energy $\rho_{\rm V}$ is converted not only into $\rho_{\rm DR}$, but also into bubble motion and gravitational waves (GWs)~\cite{Espinosa:2010hh}.
However, we may regard $\rho_{\rm DR}$ as the total hidden radiative energy component, including the relativistic bubble-wall kinetic energy and the GW energy. 
Thus, Eq.~\eqref{eq:Neff_Tn} provides a plausible measure of the hidden relativistic energy component. 
We take into account the constraint from the $\Delta N_{\rm eff}$ measurement.

\section{Super-critical PBH formation}
Here we briefly explain super-critical PBH formation. 
Since the dark sector remains subdominant, the evolution of the background density is approximately the same as in standard cosmology.
During the PT, the vacuum component $\rho_V$ evolves as $\rho_{\rm V}(t)=F(t)\Delta V$, where $F(t)$ is the volume fraction of false vacua remaining at time $t$~\cite{Liu:2021svg}. 

However, in the FVDs, the vacuum energy $\rho_{\rm V}=\Delta V$ is stored for a long time. 
Since all radiation components dilute with time while the vacuum energy remains constant, 
a large density contrast $\delta$ develops between the FVDs and the background. Practically, PBH formation is often discussed in terms of a density threshold $\delta_c$
~\cite{Carr:1975qj}.
However, the applicability of this density-fluctuation framework to super-horizon configurations remains under debate~\cite{Franciolini:2025ztf,Zou:2025sow,Wang:2026zvz,Carr:2026hot}.

In this work, we adopt the super-critical PBH formation criterion, under which PBHs can form if the radius of an FVD exceeds a critical value~\cite{Garriga:2015fdk,Deng:2017uwc,Deng:2020mds,Gouttenoire:2023gbn,Hashino:2025fse,Ning:2026nfs}.
Such objects contain a wormhole structure connecting our universe to a baby universe with eternal local inflation, and their formation cannot be captured by the standard density-fluctuation formalism.
As shown in Refs.~\cite{Garriga:2015fdk,Deng:2017uwc,Hashino:2025fse,Ning:2026nfs}, the feasibility of this mechanism can be characterized by comparing two time scales: the vacuum-energy time $t_{\rm V}$ and the horizon-crossing time $t_{\rm hc}$.  
The vacuum-energy time is defined by 
\begin{align}
t_{\rm V} = \frac{1}{2} \sqrt{ 3 M_{\rm Pl}^2/ \rho_{\rm V}} \,, 
\end{align}
where $M_{\rm Pl}$ is the reduced Planck scale. 
On the other hand, the horizon-crossing time in a radiation-dominated universe is defined by~\cite{Garriga:2015fdk, Hashino:2025fse}
\begin{align}\label{eq:tPBH}
t_{\rm hc} = H(t_{\rm in}) R^2(t_{\rm in})/2 \,, 
\end{align}
where $H(t_{\rm in})$ is the Hubble parameter at time $t_{\rm in}$, when the collapse of the FVD with initial radius $R(t_{\rm in}) \geq H^{-1}(t_{\rm in})$ starts. 
In our analysis, we define $t_{\rm in}$ by $F(t_{\rm in}) = 0.1$, meaning that the PT is completed in $90\%$ of the Universe. 
Then, the criterion for super-critical PBH formation is given by~\cite{Garriga:2015fdk, Gouttenoire:2023gbn, Hashino:2025fse, Ning:2026nfs}
\begin{align}
\label{eq:tHtV}
t_{\rm hc} > t_{\rm V} \,. 
\end{align}
If Eq.~\eqref{eq:tHtV} is satisfied, the wormhole connecting the parent and baby universes pinches off at $t = t_{\rm hc}$. 
We regard this time as the typical PBH formation time: $t_{\rm PBH} = t_{\rm hc}$. 
The validity of this expectation has been numerically confirmed~\cite{Garriga:2015fdk, Gouttenoire:2023gbn, Hashino:2025fse, Ning:2026nfs}. 
We note that Eq.~\eqref{eq:tHtV} is applicable when most regions of the Universe experience the PT, and we adopt it in our analysis. 
Typically, we obtain the approximation $t_{\rm PBH} = t_{\rm V} \simeq t_n\sqrt{\rho_{{\rm R},n}/\Delta V}$ with $t_n=1/(2H_n)$.

In determining the PBH fraction, the probability that the PT does not occur in the past light cone of a FVD is crucial. 
For a homogeneous nucleation rate, we denote this survival probability by $\mathcal{P}_{\rm surv}(t)$
for $t\geq t_{n}$. It is determined by $\Gamma(t)$ and the comoving volume $V(t',t_{\rm in})$ of the past-light-cone-bounded FVD with physical radius $R(t_{\rm in})$~\cite{Guth:1981uk}. For $V(t',t_{\rm in})$, the conventional expression is used~\cite{Gouttenoire:2023naa}
\begin{equation}
\label{eq:FVD-volumn}
V(t',t_{\rm in}) = \frac{4\pi}{3}\left[\frac{R(t_{\rm in})}{a(t_{\rm in})} + r(t_{\rm in},t') \right]^3 \equiv \frac{4\pi}{3}  \left(\frac{R_{\rm FVD}(t')}{a(t')}\right)^3 \,, 
\end{equation}
where $r(t_{\rm in},t')=\int_{t'}^{t_{\rm in}} dt'' v_w/a(t'')$ represents the comoving bubble radius. 
The probability that no PT occurs in the FVD until $t_{\rm PBH}$ is given by $P_{\rm PBH}\equiv \mathcal{P}_{\rm surv}(t_{\rm PBH})$. 
Then, the current PBH fraction $f_{\rm PBH}$ normalized to the dark matter abundance is given by (e.g., see Refs.\,\cite{Hashino:2021qoq, Gouttenoire:2023naa})
\begin{equation}
\begin{aligned}
f_{\rm PBH}&\simeq P_{\rm PBH} \frac{M_{\rm PBH} \mathcal{N}_{\rm patches}}{\rho_{\rm DM,0}\frac{4\pi}{3}H_0^{-3}}\\
\end{aligned}
\label{eq:fpbh}
\end{equation}
where $\rho_{\rm DM,0}$ is the current dark matter density, and $\mathcal{N}_{\rm patches}$ is the number of collapsing FVDs within the past light cone of the current horizon volume $\frac{4\pi}{3}H_{0}^{-3}$. 
For the PBH mass $M_{\rm PBH}$, it is approximated by the total volume energy in the FVD~\cite{Hashino:2025fse}
\begin{align}
M_{\rm PBH}&\simeq \frac{4\pi}{3} \rho_{\rm tot}(t_{\rm in}) R_{\rm FVD}^3(t_{\rm in}) \,,
\label{eq:MPBH}
\end{align}
which is the same as $M_{\rm PBH}\simeq \frac{4\pi}{3} \rho_{\rm tot}(t_{\rm PBH}) R_{\rm FVD}^3(t_{\rm PBH})$ due to energy conservation of FVDs.

We note that PBH formation via stochastic FOPTs at low temperatures is ruled out by a core tension: (1) a weak and rapid FOPT is required to evade the $\Delta N_{\rm eff}$ constraint~\cite{Bai:2021ibt, An:2026hiq}, and (2) a slow FOPT is needed to obtain long-lived FVDs~\cite{An:2026hiq}. 
We have numerically confirmed that PBH formation based on conventional stochastic FOPTs at low temperatures is not feasible, even if we adopt a higher-order exponential bubble nucleation rate (details are in App.~\hyperref[sec:app-higher-order]{A}).

\section{Catalyzed phase transitions by domain walls}
Since SMPBH formation via pure stochastic FOPTs is challenging, we discuss bubble nucleation catalyzed by impurities and show that SMPBHs can form while satisfying the above experimental constraints.  

Early-universe impurities, such as topological defects or PBHs, are naturally produced in many scenarios beyond the SM and can act as nucleation seeds through catalysis. The impact of such catalysts, including PBHs~\cite{Hiscock:1987hn, Berezin:1990qs, Gregory:2013hja, Burda:2015yfa, Mukaida:2017bgd, Kohri:2017ybt, Oshita:2019jan, Gregory:2020cvy, Hayashi:2020ocn,  El-Menoufi:2020ron, Shkerin:2021zbf, Shkerin:2021rhy, Strumia:2022jil, Briaud:2022few, Jinno:2023vnr, Zhong:2025xwm, Chang:2025rda, Rossi:2025fix, Tanaka:2026geo}, monopoles~\cite{Steinhardt:1981ec, Kumar:2009pr, Kumar:2010mv, Agrawal:2022hnf}, and DWs~\cite{Blasi:2022woz, Blasi:2023rqi, Agrawal:2023cgp, Sassi:2025dyj,Ge:2023rrq}, has been extensively studied. Following Refs.~\cite{Blasi:2022woz, Blasi:2023rqi}, we investigate the case where DWs catalyze dark-sector PTs. 

For the bubble nucleation rate around DWs per unit surface area and unit time, we have~\cite{Blasi:2022woz}
\begin{align}
\label{eq:gamma_dw}
\gamma_{\rm DW} \simeq \sigma_{\rm DW} e^{- S_{\rm inh}} \,,
\end{align}
where $\sigma_{\rm DW}$ and $S_{\rm inh}$ represent the surface energy density of DWs and the bounce action around a DW, respectively. In the following, we consider the case in which bubble nucleation occurs only through DW catalysis rather than through the usual stochastic bubble nucleation, as shown in Fig.~\ref{fig:schematic-diagram}. 

If there are bulk regions without DWs, bubble nucleation does not occur there.
For simplicity, we assume that such void regions are statistically rare.
This means that every Hubble patch contains at least part of a DW on average, and that DWs play an essential role in completing the dark-sector PT.
In addition, we assume that DWs are distributed homogeneously throughout the Universe with number density $n_{\rm DW}$. 
Based on these assumptions, the bubble nucleation rate per unit volume and time due to DWs is expressed as
\begin{align}
\Gamma_{\rm DW}(t) = n_{\rm DW}(t) L^2 \gamma_{\rm DW} \,,
\end{align}
where $L$ denotes the typical length of DWs. 

For the bounce action $S_{\rm inh}$ in Eq.~\eqref{eq:gamma_dw}, we assume
\begin{align}
S_{\rm inh} \sim -\frac{\zeta_{\rm DW}^2}{2}(t - t_{n})^2 \,. 
\end{align}
If $\zeta_{\rm DW}$ is sufficiently large, bubble nucleation occurs suddenly at $t = t_{n}$ around DWs.
Thus, the bubble nucleation rate is approximately
\begin{align}
\label{eq:GammaDW_Cn}
\Gamma_{\rm DW} 
\sim \frac{\sqrt{2 \pi} L^2 \sigma_{\rm DW}}{\zeta_{\rm DW}} n_{\rm DW}  \delta (t - t_{n}) \equiv C_{\rm DW} n_{\rm DW} \delta(t - t_{n}) \,, 
\end{align}
where $C_{\rm DW} \equiv \sqrt{2 \pi} L^2 \sigma_{\rm DW}/\zeta_{\rm DW}$ and $n_{\rm DW} \equiv n_{\rm DW}(t_{n})$.
Eq.~\eqref{eq:GammaDW_Cn} contains the dimensionless parameter $C_{\rm DW}$, which is absent for point-like catalysts~\cite{Jinno:2023vnr, Zhong:2025xwm}.
This additional parameter is important for catalyzed FOPT dynamics.
A large $C_{\rm DW}$ is naturally realized when the DWs possess a large $\sigma_{\rm DW}$~\cite{Blasi:2022woz}. 

\begin{figure}[t]
\centering
\includegraphics[width=0.98\linewidth]{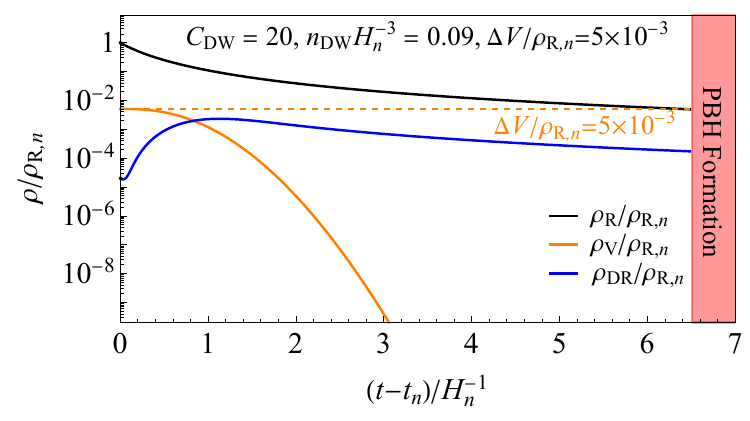}
\caption{
Time evolution of the energy densities for FOPTs catalyzed by DWs for benchmark A in Eq.~\eqref{eq:BP_A}. 
Each energy density is normalized to the initial SM radiation energy density $\rho_{{\rm R},n}$. 
In the red-shaded region, PBH formation has been completed in the FVDs when $\Delta V\gtrsim \rho_{{\rm R}}(t_{\rm PBH})$.}
\label{fig:density_t}
\end{figure}

Employing Eq.~\eqref{eq:GammaDW_Cn}, the averaged false vacuum fraction is given by~\cite{Jinno:2023vnr}
\begin{align}
\label{eq:Ft_catalyst}
F(t) = \exp \left[ - \frac{4\pi}{3} C_{\rm DW} n_{\rm DW} r^3(t, t_{n}) \right]  \quad (t > t_{n} ) \,, 
\end{align}
where $a(t_{n}) =1$ is assumed. 
Using Eq.~\eqref{eq:Ft_catalyst}, the time dependence of the vacuum energy is approximated by $\rho_{\rm V} = F(t) \Delta V$. 
In Fig.~\ref{fig:density_t}, the time evolution of each energy component is shown for benchmark A:
\begin{align}
\label{eq:BP_A}
C_{\rm DW} = 20 \,, ~\Delta V/\rho_{{\rm R},n} = 5 \times 10^{-3} \,, ~ L H_{n} = 7 \,,
\end{align}
with $n_{\rm DW} H_{n}^{-3} = 0.09$. The required density and length scale of DWs can be realized in scenarios where the DW production occurs during inflation and DWs are subsequently diluted~\cite{Harigaya:2022pjd}.
The vacuum energy is efficiently converted into DR through the catalytic effect of DWs within a few Hubble times, while in the red shaded region the long-lived FVDs satisfying Eq.~\eqref{eq:tHtV} collapse into super-critical PBHs.

For the survival probability of a FVD, a careful derivation is needed. 
The survival probability in our setup includes two contributions: (1) how many DWs penetrate the FVD, and (2) what is the probability that each DW inside the FVD does not nucleate vacuum bubbles? 
Supposing that the surface area of the $i$-th DW penetrating the FVD is given by $S_{i}$, the survival probability is
\begin{align}
\label{eq:Psurv_sum}
\mathcal{P}_{\rm surv}(t) = \sum_{k = 0}^{N_{\rm DW}} \mathcal{P}_{k} \int \left( \prod_{i=1}^{k} d S_{i} \, p(S_{i}) \, e^{- \int dt \gamma_{\rm DW}(t) S_{i}(t)} \right) \,, 
\end{align}
where $\mathcal{P}_{k}$ is the probability that $k$ DWs \textit{penetrate the FVD}, and $N_{\rm DW}$ is the total number of DWs in the whole Universe. 
We assume that $\mathcal{P}_{k}$ obeys a Poisson distribution: $\mathcal{P}_{k} = \braket{n}^{k} e^{-\braket{n}}/k!~(k = 0\,,1\,,2\,,\cdots, N_{\rm DW})$, where $\braket{n}$ represents the average number of DWs \textit{penetrating the FVD}. 
When DWs are distributed homogeneously, we obtain $\braket{n} \simeq 2 n_{\rm DW} L^2 R_{\rm FVD}$, with $R_{\rm FVD}=R_{\rm FVD}(t_n)$. 
The function $p(S_{i})$ is the distribution function of the surface area $S_i$ within the FVD for the $i$-th DW. 
When $L \gg R_{\rm FVD}$, $p(S_{i})$ takes a simple form. 
The derivation of $p(S_{i})$ and $\braket{n}$ is given in App. \hyperref[sec:p(S)derivation]{B}.  
Performing the summation and integral in Eq.~\eqref{eq:Psurv_sum} in the large $N_{\rm DW}$ limit, we obtain
\begin{align}
\label{eq:Psurv_catalyst}
    \mathcal{P}_{\rm surv}^{\rm DW}(t)\simeq\exp \left[ - \braket{n} \left( 1 - \frac{F_{\rm D}(\sqrt{K_{\rm DW}})}{\sqrt{K_{\rm DW}}}  \right) \right] \,,
\end{align}
with $K_{\rm DW} \equiv \pi C_{\rm DW} R_{\rm FVD}^2 /L^2$. 
The function $F_{\rm D}(x)$ denotes the Dawson integral. 
The probability that no PT occurs in the FVD until $t_{\rm PBH}$ is given by $P_{\rm PBH}\equiv \mathcal{P}^{\rm DW}_{\rm surv}(t_{\rm PBH})$.
Notably, since $\braket{n} \propto R_{\rm FVD}$, the scaling of $R_{\rm FVD}$ in $\mathcal{P}_{\rm surv}$ in our scenario is $\exp \left(-R_{\rm FVD} \right)$, which is milder than the usual scaling $\exp \left(-R_{\rm FVD}^3 \right)$.

\begin{figure}[t]
\centering
\includegraphics[width=0.98\linewidth]{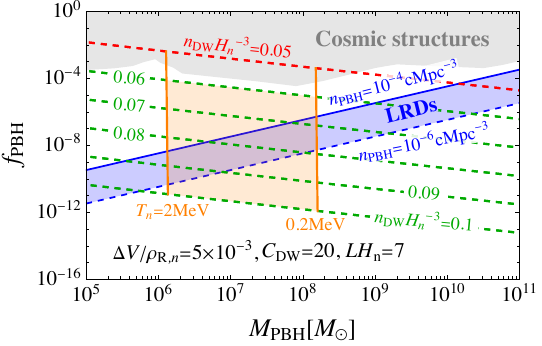}
\caption{
Prediction for the PBH mass and fraction with different DW densities (green dashed lines) for benchmark A.
The orange band indicates $T_n \in [0.2\,{\rm MeV},\,2\,{\rm MeV}]$, which yields $\mathcal{O}(10^{6}-10^{8})\,M_\odot$ PBHs relevant to LRD simulations in Ref.~\cite{Zhang:2025oyl}. 
Above the red line, corresponding to $n_{\rm DW} H_n^{-3} \leq 0.05$, the PT completion condition in Eq.~\eqref{eq:Nbubble_DW} cannot be satisfied. 
The blue region shows the LRD-favored PBH parameter space~\cite{DeLuca:2025nao}, 
while the gray region is excluded by existing constraints from cosmic structures~\cite{Carr:2018rid, Carr:2020erq}. 
For benchmark A, the cosmological constraints do not visibly cut into the parameter space, but become relevant for larger $\Delta V$ and smaller $T_n$, as shown in App.~\hyperref[sec:details-plots]{C}.
}
\label{fig:fpbh_mpbh_dw}
\end{figure}

In order to complete the PT, the number of nucleated bubbles $N_{\rm bubble}$ in each Hubble patch should exceed unity. 
This PT completion requirement is formulated as
\begin{align}
\label{eq:Nbubble_DW}
N_{\rm bubble} = \int_{t_{n}}^{\infty} dt \frac{\Gamma_{\rm DW}}{H^3} = C_{\rm DW} n_{\rm DW} H_{n}^{-3} \geq 1\,. 
\end{align}
Physically, $C_{\rm DW}$ roughly corresponds to the ratio of $N_{\rm bubble}$ to the number of DWs in a Hubble patch, characterizing the strength of the catalytic effect.

In Fig.~\ref{fig:fpbh_mpbh_dw}, the predicted PBH mass and fraction are shown for different $n_{\rm DW}$ (dashed green lines) in benchmark A of Eq.~\eqref{eq:BP_A}. 
For the nucleation temperature $T_{n}$, we take $T_{n} \in [0.2\,{\rm MeV}, 2\,{\rm MeV}]$. 
The left (right) orange solid line in Fig.~\ref{fig:fpbh_mpbh_dw} corresponds to $T_{n} = 2\,{\rm MeV}~(0.2\,{\rm MeV})$. 
Above the red solid line, the PT completion condition in Eq.~\eqref{eq:Nbubble_DW} cannot be satisfied. As in many other PT scenarios, the mass of PBHs produced here will be nearly monochromatic. The figure shows that $M_{\rm PBH}$ is mainly determined by $T_n$ through the FVD radius $R(t_{\rm in})$, while $f_{\rm PBH}$ is mainly controlled by $n_{\rm DW}$. With smaller $T_n$, our mechanism can produce SMBH seeds as massive as $10^{10}~M_{\odot}$ with a sufficient number density to potentially account for the entire LRD population.

We have confirmed numerically that the constraints from FOPT completion, curvature perturbations, and 
$\Delta N_{\rm eff}$ are satisfied for the parameter region shown in Fig.~\ref{fig:fpbh_mpbh_dw}.
The reason is that DWs drive rapid FOPT completion through their strong catalytic effect, while FVDs survive long enough primarily because of the rare statistics of the DW distribution. 
Thus, our scenario does not suffer from the core tension of conventional stochastic FOPTs at low scales. The cosmological constraints become visible for larger $\Delta V$ and small $T_n$, as shown in App.~\hyperref[sec:details-plots]{C}.

\begin{figure}[t]
    \centering
    \includegraphics[width=0.98\linewidth]{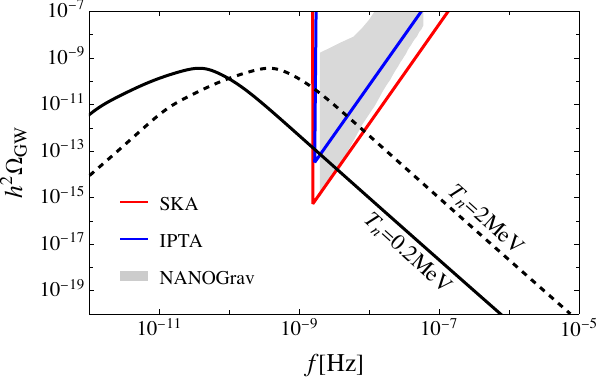}
    \caption{Prediction for the GW spectrum for two $T_{n}$ values with $n_{\rm DW}H_{n}^{-3} = 0.09$ in benchmark A. For the bubble-wall velocity, we take $v_{w} = 0.95$. The red and blue solid lines are the expected sensitivity curves of SKA~\cite{Carilli:2004nx} and IPTA~\cite{Hobbs:2009yy}. The gray shaded region indicates the signal observed by NANOGrav~\cite{NANOGrav:2023gor, NANOGrav:2023hvm}. }
    \label{fig:GWplot}
\end{figure}

Furthermore, SGWBs can be generated during the catalyzed dark-sector PT. 
We use the fitting functions for the GW spectrum analysis summarized in Ref.~\cite{Caprini:2024hue}. 
In Fig.~\ref{fig:GWplot}, we focus on benchmark A with $n_{\rm DW} H_{\rm n}^{-3} = 0.09$ and two different $T_n$. We find that the SGWB signal can be partially tested by current and future PTA experiments such as NANOGrav~\cite{NANOGrav:2023gor}, SKA~\cite{Carilli:2004nx}, IPTA~\cite{Hobbs:2009yy}, and CPTA~\cite{Xu:2023wog},  see also the discussion in App.~\hyperref[sec:app-gw-signal]{D}.

\section{Conclusions and Discussions}
We proposed a new origin for the SMBHs associated with JWST LRDs. We considered a general \emph{catalyzed} FOPT in a subdominant, decoupled dark sector at sub-MeV scales, with cosmic expansion always governed by SM radiation and the latent heat deposited solely into DR. 
We showed that, when the dark-sector PT is catalyzed by DWs, the formation of SMPBHs with masses up to $M_{\rm PBH}\sim 10^{10} M_\odot$ and an abundance appropriate for LRDs is feasible.
We also confirmed that strong cosmological and phenomenological constraints, including CMB spectral distortions, $\Delta N_{\rm eff}$, PT completion, and DW overclosure, can be consistently satisfied in this scenario. 
Such SMPBHs can naturally serve as seeds of compact, low-metallicity, BH-dominated galaxies associated with LRDs.

At the same time, the dark PT generates SGWBs peaking at $f_{\rm peak}\sim \mathcal{O}(10^{-2} \text{--} 10^{-1})\,\mathrm{nHz}$, partly accessible to current and future PTAs. 
A joint analysis of the LRD number density, $\Delta N_{\rm eff}$ constraints, and GW data therefore provides a concrete way to test this sub-MeV catalyzed dark-sector PT origin of SMBH seeds. In this sense, the discovery of LRDs may offer an indirect hint of a sub-MeV-scale FOPT and the presence of DWs.

\section*{Acknowledgements}
We thank Xuejian Shen for useful discussions. The work of J.G is supported by the Postdoctoral Fellowship Program (Grade C) of China Postdoctoral Science Foundation under Grant No. GZC20252775.
The work of J.L. is supported by the National Science Foundation of China under Grant No. 12235001, No. 12475103 and State Key Laboratory of Nuclear Physics and Technology under Grant No. NPT2025ZX11.
The work of X.P.W. is supported by National Science Foundation of China under Grant No. 12375095, and the Fundamental Research Funds for the Central Universities.  HX is supported by the U.S. Department of Energy under grant
DE-SC0026297.
J.L. and X.P.W. also thank the Mainz Institute for Theoretical Physics (MITP) of the PRISMA+ Cluster of Excellence (Project ID 390831469) for its hospitality and partial support during the completion of this work. 
The authors gratefully acknowledge the valuable discussions and insights provided by the members of the Collaboration of Precision Testing and New Physics.

\clearpage


\appendix

\begin{center}
{\large \bf Appendix }
\label{sec:app}
\end{center}

This appendix provides detailed derivations, extended calculations, and mathematical proofs supporting the results presented in the main text.

\section{Higher-order nucleation rate}
\label{sec:app-higher-order}

In order to describe model-independent properties of FOPTs, the bubble nucleation rate is often approximated by~\cite{Megevand:2016lpr}
\begin{equation}
\label{eq:Gamma_quadratic}
    \Gamma (t) \simeq H_n^4 e^{\beta(t-t_n)-\frac{\zeta^2}{2}(t-t_n)^2} = H_n^4 e^{\beta(t-t_n)\left(1-\frac{t-t_n}{2t_s}\right)} \,, 
\end{equation}
where $\beta$ and $\zeta$ correspond to the first- and second-order coefficients of the Taylor-expansion for the bounce action around $t_n$. We introduce $t_s=\beta/\zeta^2$ for convenience. The quadratic expansion of the bounce action allows for the possibility of surviving the FVD after $t_n$.

\begin{figure}[tbh]
\centering
\includegraphics[width=0.9\linewidth]{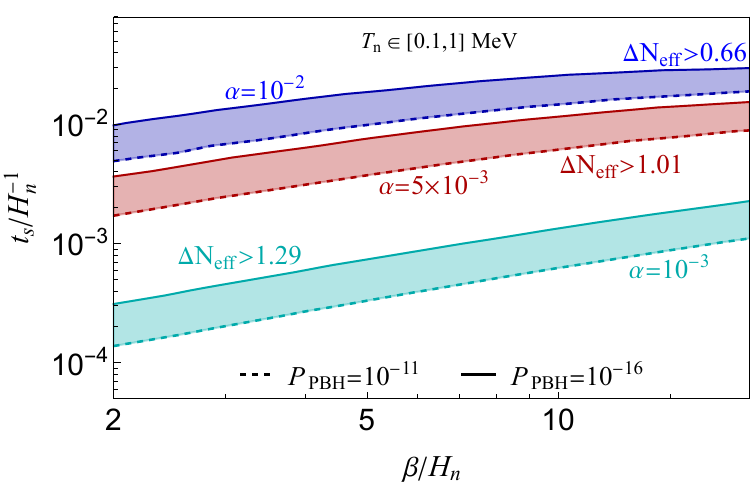}
    \caption{
    PBH formation probability $P_{\rm PBH}$ contours plots as functions of $\beta/H_n$ and $t_s/H_n^{-1}$ for each $\alpha$ value.
    We take $\rho_{{\rm DR},n}/\rho_{{\rm R},n} = 2\times 10^{-5}$ in this plot.
    }
    \label{fig:ts_beta}
\end{figure}

With this nucleation rate, by numerically solving the energy density evolutions and considering the PBH formation super-critical requirement in Eq.~\eqref{eq:tHtV}, we can obtain the suitable parameter spaces for LRDs.
In Fig.~\ref{fig:ts_beta}, the parameter dependence of $P_{\rm PBH}$ is shown for each $\alpha$ value. 
Each color band corresponds to the region with $10^{-16} < P_{\rm PBH} < 10^{-11}$ where the SMPBHs possess suitable fraction for the LRD phenomenology~\cite{Shen:2025evo, DeLuca:2025nao}. However, the tiny $t_s$ usually means a slow PT, which results in a larger PT completion time $t_{\rm PT}$. As a result, during this epoch, the radiation energy density is considerably diluted, resulting in a large ratio $\rho_{\rm DR}/\rho_{\rm R}$ after the PT. This is in tension with the BBN constraints on $\Delta N_{\rm eff}\lesssim 0.3$, as shown in this figure. 
The same results have been confirmed in Ref.~\cite{An:2026hiq} in a case without the quadratic term in Eq.~\eqref{eq:Gamma_quadratic}. 

We have checked that the conventional FOPT always suffers from the following tension: on the one hand, BBN constraints typically require the PT to proceed very rapidly, so that the ratio $\rho_{\rm DR}/\rho_{\rm R}$ remains small and evades these constraints. On the other hand, a sufficiently long-lived FVD usually requires the PT to occur more gradually, in order to realize a sizable survival probability $\mathcal{P}_{\rm surv}$. We have confirmed that this situation becomes even more pronounced for the linearly expanded form $\Gamma (t)=H_n^4 e^{\beta(t-t_n)}$. 
This tension is not prominent in the extensively studied supercooled PT scenario \cite{Lewicki:2023ioy,Kawana:2022olo,Gouttenoire:2023naa}, because in that case the second term in Eq.~\eqref{eq:FVD-volumn}, which characterizes the past light cone part of the FVD $r(t,t')$, is extremely small and can be neglected. In our mechanism, by contrast, this term is comparable to the radius of FVD, which strongly suppresses the survival probability $\mathcal{P}_{\rm surv}$.

\section{Derivations for the survival probability } \label{sec:p(S)derivation}

In this sub-appendix, the distribution function $p(S_{i})$ and the average number of DWs penetrating the FVD $\braket{n}$ introduced in Eq.~\eqref{eq:Psurv_sum} are derived in order. 

\begin{figure}[tbh]
    \centering
    \includegraphics[width=0.8\linewidth]{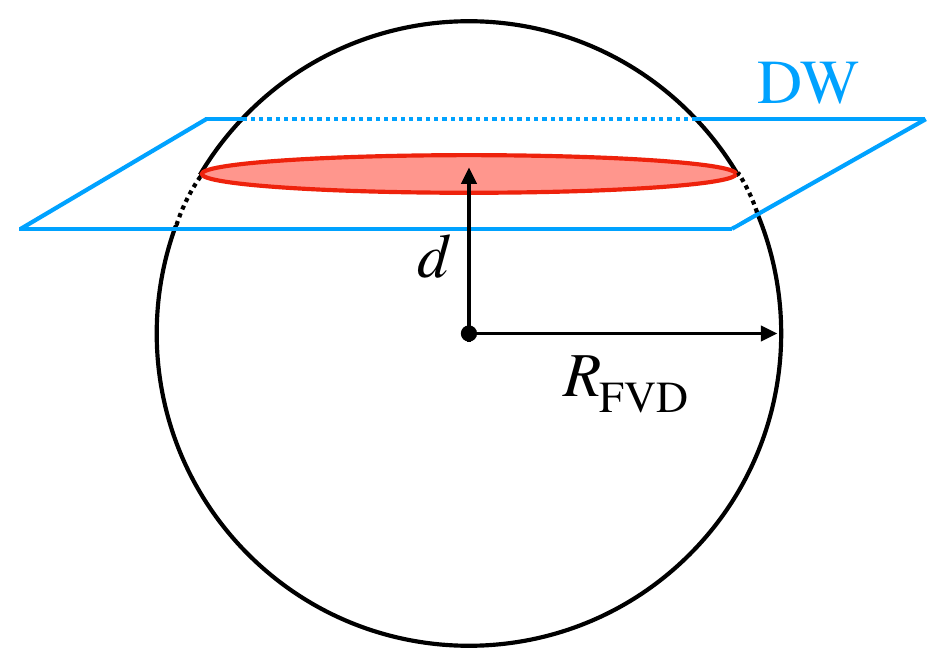}
    \caption{
    Conceptual figure for the system with the DW penetrating the FVD with the radius $R_{\rm FVD}$ at the point $d$.
    The red region denotes the surface area of the DW within the FVD. 
    }
    \label{fig:sphere}
\end{figure}

We first derive the distribution function $p(S_{i})$. 
In order to derive it, we have to specify a system in which we are interested. 
In our analysis, it is assumed that the DW size is larger than the radius of the FVD: $L \gg R_{\rm FVD}$. 
In this case, we can expect that the DW penetrating the FVD is regarded as a plane, as shown in Fig. \ref{fig:sphere}. 
Since DWs are distributed homogeneously in the Universe in our setup, the probability that a DW penetrates the FVD at the point $d$ is uniform. 
Therefore, the corresponding distribution function $\tilde{p}(d)$ is given by 
\begin{align}
\tilde{p}(d) = \frac{1}{R_{\rm FVD}} \quad \text{with} \quad \int_{0}^{R_{\rm FVD}} dx \tilde{p}(x) = 1 \,. 
\end{align}
Then, the distribution function $p(S)$ of the surface area $S$ for the red region in Fig.~\ref{fig:sphere} obeys 
\begin{align}
\label{eq:p(S)_app}
p(S) = \tilde{p}(x) \left| \frac{dx}{dS} \right| = \frac{1}{2 \pi R_{\rm FVD} \sqrt{R_{\rm FVD}^2 - S/\pi}} \,,  
\end{align}
with 
\begin{align}
\int_{0}^{\pi R_{\rm FVD}^2} p(S) dS = 1 \,. 
\end{align}
In our analysis, we adapt Eq.~\eqref{eq:p(S)_app} to estimate the survival probability in Eq.~\eqref{eq:Psurv_sum}. 

In the above discussion, we have considered the system in which one DW intersects with the FVD. 
However, in general, multiple DWs can exist in the FVD. 
In such a case, the probability that there is no PT in the FVD is given by Eq.~\eqref{eq:Psurv_sum}. 
Substituting Eq.~\eqref{eq:p(S)_app} into Eq.~\eqref{eq:Psurv_sum} and performing simple calculations, we obtain 
\begin{align}
\mathcal{P}_{\rm surv}^{\rm DW}(t) = \sum_{k = 0}^{N_{\rm DW}} \frac{ \braket{n}^{k} e^{- \braket{n}} }{k!} \left( \frac{F_{\rm D}(\sqrt{K_{\rm DW}})}{\sqrt{K_{\rm DW}}} \right)^{k} \,,
\end{align}
with $K_{\rm DW} = \pi C_{\rm DW} R_{\rm FVD}^2 /L^2$.
Since we have assumed that every Hubble patch includes a part of DWs, the total DW number $N_{\rm DW}$ in the whole universe should be large. 
In the limit $N_{\rm DW} \gg 1$, $\mathcal{P}_{\rm surv}(t)$ takes
\begin{align}
\label{eq:Psurv_n_app}
\mathcal{P}_{\rm surv}^{\rm DW}(t) \simeq \exp \left[ - \braket{n} \left( 1 - \frac{F_{\rm D}(\sqrt{K_{\rm DW}})}{\sqrt{K_{\rm DW}}} \right)  \right] \,, 
\end{align}
which is the same as Eq.~\eqref{eq:Psurv_catalyst}. 

\begin{figure}[tbh]
    \centering
    \includegraphics[width=0.8\linewidth]{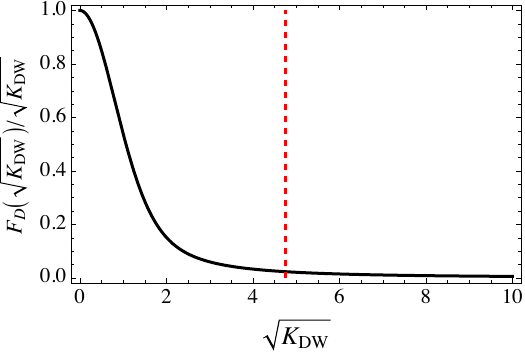}
    \caption{
    The form of $F_{D}(\sqrt{K_{\rm DW}})/\sqrt{K_{\rm DW}}$. 
    The red dashed vertical line is the value of $\sqrt{K_{\rm DW}}$ in the case with $n_{\rm DW} H_{n}^{-3} = 0.09$ and $T_{n} = 2\,{\rm MeV}$ in Fig.~\ref{fig:fpbh_mpbh_dw}. 
    }
    \label{fig:xDawson}
\end{figure}

In Fig.~\ref{fig:xDawson}, the form of the function $F_{D}(x)/x$ is shown. 
This figure indicates that the function $F_{D}(x)/x$ becomes unity if $x \to 0$. 
This implies that $\mathcal{P}_{\rm surv}^{\rm DW}(t)$ given in Eq.~\eqref{eq:Psurv_n_app} becomes unity if $K_{\rm DW} \to 0$, which corresponds to the case where the catalysis effect of DWs is negligible. 
As a result, the survival probability and the corresponding PBH abundance can be large. 

On the other hand, if $K_{\rm DW} \gg 1$, $\mathcal{P}_{\rm surv}^{\rm DW}(t)$ takes the following simple form 
\begin{align}
\label{eq:Psurv_largeKDW}
\mathcal{P}_{\rm surv}^{\rm DW}(t) \simeq e^{- \braket{n}} \,. 
\end{align}
This result implies that a large $\mathcal{P}_{\rm surv}^{\rm DW}$ requires a rare situation where there is no DW in the FVD to avoid the significant catalysis effect. 
For instance, as shown in Fig.~\ref{fig:xDawson}, we obtain $\sqrt{K_{\rm DW}} = 4.76$ in the case with $n_{\rm DW} H_{n}^{-3} = 0.09$ and $T_{n} = 2\,{\rm MeV}$ in Fig.~\ref{fig:fpbh_mpbh_dw}. 
Therefore, in such a case, the survival probability is almost determined by Eq.~\eqref{eq:Psurv_largeKDW} and it implies that the smallness of the PBH fraction is mainly determined by the probability that no DW exists in the FVD.

We finally derive the expression for $\braket{n}$. 
Since we assume that the DWs are distributed homogeneously with the number density $n_{\rm DW}$, the expectation value of the total surface area of DWs in the sphere $R_{\rm FVD}$ is given by 
\begin{align}
\braket{S} \simeq \frac{4}{3} \pi n_{\rm DW}   R_{\rm FVD}^3 \times L^2 \,,
\label{eq:bigS-ave}
\end{align}
where the first part represents the expected number of DW in the sphere $R_{\rm FVD}$. For the case with DWs with a sub-horizon size ($R_{\rm FVD} > L$), the entire DW is contained in the sphere $R_{\rm FVD}$ with volume $V_{\rm FVD}= 4\pi R_{\rm FVD}^3/3$, so that $\braket{S}$ is exactly the number of DWs times $L^2$. On the other hand, when the size of DWs is super-horizon ($R_{\rm FVD} < L$), we can rewrite $n_{\rm DW} \equiv N_{\rm DW}/V_{\rm uni}$, where $V_{\rm uni}$ represents the whole universe volume. By reforming Eq.~\eqref{eq:bigS-ave}, we have
\begin{align}
\frac{\braket{S}}{N_{\rm DW}L^2} = \frac{V_{\rm FVD}}{V_{\rm uni}},
\end{align}
where $N_{\rm DW}L^2$ is the total surface area of DWs in the universe. Therefore, the above formula is exactly the consequence of the homogeneous assumption for the DW distribution.

On the other hand, under the condition that one DW intersects with the sphere $R_{\rm FVD}$ as shown in Fig.~\ref{fig:sphere}, the expectation value of the surface area in the FVD sphere (the red region) is given by 
\begin{align}
\braket{s} = \int_{0}^{\pi R_{\rm FVD}^2} S p(S) dS = \frac{2}{3} \pi R_{\rm FVD}^2 \,. 
\end{align}
Since $\braket{n}$ represents the average number of DWs \textit{penetrating} the FVD, it can be expressed by
\begin{align}
\label{eq:n_ave_app}
\braket{n} \simeq \frac{\braket{S}}{\braket{s}} = 2 n_{\rm DW} L^2 R_{\rm FVD} \,. 
\end{align}
We note that $\braket{n}$ does not mean the average number of DWs existing inside the FVD. 

\section{Details of Cosmological Constraints on Catalyzed Phase Transitions} \label{sec:details-plots}

This section elaborates on cosmological constraints imposed on catalyzed cosmological phase transitions. In general, the latent heat, phase transition duration, and bubble nucleation time collectively determine the final $\Delta N_{\rm eff}$ and $\mu$-type spectral distortion.

\begin{figure}[htbp]
    \centering
    \includegraphics[width=0.9\linewidth]{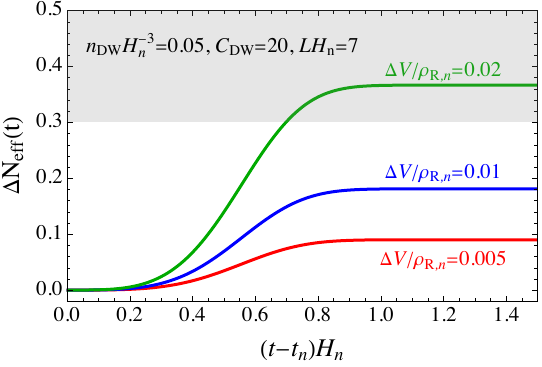}
    \includegraphics[width=0.95\linewidth]{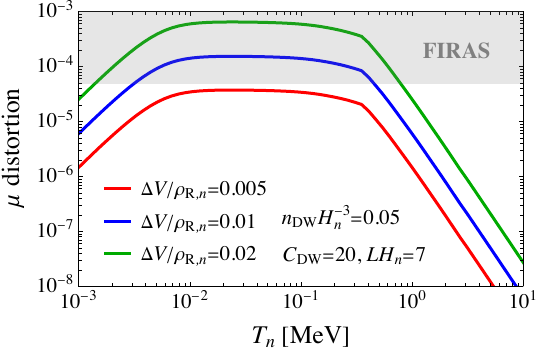}
    \caption{Cosmological observables for catalyzed phase transitions. (Top) Evolution of $\Delta N_{\text{eff}}$ for $n_{\text{DW}} H_n^{-3}=0.05$, $C_{\text{DW}}=20$, and $LH_n=7$ across different latent heat fractions. (Bottom) Corresponding $\mu$-distortion as a function of the nucleation temperature $T_n$.}
    \label{fig:BBN_CMB}
\end{figure}

The top panel of Fig.~\ref{fig:BBN_CMB} illustrates the temporal evolution of $\Delta N_{\text{eff}}$ for a benchmark parameter set ($n_{\text{DW}}H_n^{-3}=0.05, C_{\text{DW}}=20, LH_n=7$). As the phase transition progresses, $\Delta N_{\rm eff}$ increases monotonically before saturating at an asymptotic value. Typically, a higher latent heat fraction $\Delta V/\rho_{\text{R},n}$ leads to a larger final $\Delta N_{\rm eff}$. For this specific parameter combination, the current BBN and CMB constrain $\Delta V/\rho_{\text{R},n} \gtrsim 0.01$. In particular, the three setups shown reach their maximum value $\Delta N_{\rm eff}$ almost simultaneously. This behavior arises because the effective duration $\bar{\beta}$ is primarily governed by the product $n_{\rm DW} \cdot C_{\rm DW}$, as indicated by Eqs.~\eqref{eq:Ft_catalyst} and \eqref{eq:beta}.
Moreover, this figure shows that the dark-sector phase transition completes rapidly in our setup.

The lower panel of Fig.~\ref{fig:BBN_CMB} depicts the nucleation temperature dependence of the $\mu$-distortions generated by the catalyzed phase transition. We used the formalism summarized in Ref.~\cite{Greene:2026gnw}. The distortions peak when the nucleation temperature $T_n$ falls within the $0.01\text{--}0.3$ MeV range. Outside this interval, the signal is suppressed, reflecting the specific redshift sensitivity window of the $\mu$-distortion epoch [see Eq.~\eqref{eq:mu-distortion}]. When compared against FIRAS observations ($\mu < 4.7 \times 10^{-5}$), models with significant latent heat ($\Delta V/\rho_{\text{R},n} \gtrsim 0.005$) are disfavored for sub-MeV phase transitions, placing stringent bounds on the available parameter space. We have also numerically checked that the $y$-distortion mainly constrains phase transitions at $T_n\sim {\rm keV}$ and is generally weaker than the $\mu$-distortion constraint.

\begin{figure}[htbp]
    \centering
    \includegraphics[width=0.95\linewidth]{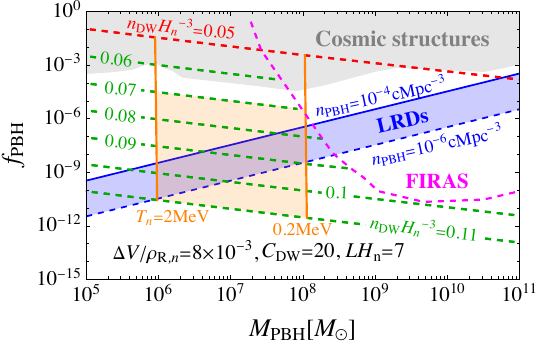}
    \caption{Constraints on the PBH mass-abundance relation ($f_{\text{PBH}}$ vs. $M_{\text{PBH}}$) for varying domain wall number densities $n_{\text{DW}}$, assuming $\Delta V/\rho_{\text{R},n} = 0.008$. The regions above the magenta-dashed curve indicate the parameter space excluded by FIRAS $\mu$-distortion observations~\cite{Fixsen:1996nj, Bianchini:2022dqh}.}
    \label{fig:fpbh_mbph_DW_008}
\end{figure}

To further assess the viability of the model, Fig.~\ref{fig:fpbh_mbph_DW_008} shows an additional benchmark set, obtained by changing $\Delta V/\rho_{{\rm R},n}$ to $8\times 10^{-3}$, which can produce PBHs massive enough to account for the observed high-redshift LRDs. 
However, for larger PBH masses, the parameter space is strongly constrained by the FIRAS $\mu$-distortion bound, which effectively disfavors smaller DW number densities $n_{\rm DW}$.
This is because a smaller $n_{\rm DW}$ implies a slower PT.
Since a slower PT generates a larger curvature power spectrum $P_{\zeta}$ at small scales, the spectral-distortion constraint becomes more stringent.

Furthermore, it is crucial to ensure that the DWs do not overclose the Universe. 
The relic abundance $f_{\rm DW} \equiv \Omega_{\rm DW} / \Omega_{\rm DM}$, representing the DW energy density normalized to the observed dark matter density, can be expressed as:
\begin{equation}
f_{\rm DW} \simeq 9.7 \times 10^{-84} \left[ \frac{T_n}{0.1 \text{ MeV}} \right]^5 \left[ \frac{n_{\rm DW} H_n^{-3}}{0.1} \right] \left[ \frac{C_{\rm DW}}{20} \right] \left[ \frac{\zeta_{\rm DW}}{10 H_n} \right] .
\end{equation}
This implies that the DW contribution to the total energy density remains safely subdominant throughout the cosmological evolution.

\section{Gravitational wave signals} \label{sec:app-gw-signal}

Next, we discuss the prediction of a SGWB produced by the dark sector FOPT. 
The properties of FOPT are determined mainly by its latent heat $\alpha = \rho_{\rm V}/\rho_{\rm R}$ and duration $\overline{\beta}$. 
For the parameter $\overline{\beta}$, we use the following definition suggested in Ref.~\cite{Jinno:2023vnr}
\begin{align}\label{eq:beta}
\overline{\beta} = \left. -\frac{d \ln F}{dt} \right|_{t = t_{\rm perc}} \,,
\end{align}
where $t_{\rm perc}$ represents the percolation time.
For the parameter $\alpha$, since the cosmological expansion is dominated by the SM radiation ($\rho_{\rm R} \gg \rho_{\rm V} \gg \rho_{\rm DR}$) in our scenario, it is small in general. 

When the FOPT occurs in the early Universe, the stochastic GW can be produced by three sources: (1) bubble collision, (2) sound wave, and (3) magnetohydrodynamic turbulence (i.e., see references in Refs.~\cite{Caprini:2015zlo, Caprini:2024hue}). 
In our analysis, we employ the fitting functions for the GW spectrum summarized in Ref.~\cite{Caprini:2024hue}. 

In Fig.~\ref{fig:GWplot}, the prediction on the GW spectra is shown with $\Delta V/\rho_{{\rm R},n} = 0.005$, $C_{\rm DW} = 20$, $\rho_{{\rm DR},n}/\rho_{{\rm R},n}=2\times 10^{-5}$, $n_{\rm DW}H_{n}^{-3} = 0.09$ and $L H_{n} = 7$. 
As shown in Fig.~\ref{fig:fpbh_mpbh_dw}, this benchmark point predicts the formation of SMPBHs whose properties well match the LRD phenomenology. 
In the temperature range $T_{n} \in [0.2\,{\rm MeV}, 2\,{\rm MeV}]$, the peak frequency is within $f_{\rm peak} \in [3.71 \times 10^{-2} \,{\rm nHz}, ~ 3.73 \times 10^{-1} \,{\rm nHz}]$. 
However, the peak GW amplitudes are hardly changed. 
These GW spectra can be partially observed at SKA~\cite{Carilli:2004nx} and IPTA~\cite{Hobbs:2009yy} in addition to NANOGrav~\cite{NANOGrav:2023gor}. 

\bibliographystyle{JHEP}
\bibliography{ref}

\end{document}